\documentclass[prl,aps,amssymb,nofootinbib,twocolumn,showpacs]{revtex4}
\usepackage{amsmath}
\usepackage{amssymb}
\usepackage{amsthm}
\usepackage{amsfonts}
\usepackage{enumerate}
\usepackage{latexsym}
\usepackage[dvips]{graphicx}

\newcommand{\beq}{\begin{equation}}
\newcommand{\eneq}{\end{equation}}
\newcommand{\beqs}{\begin{equation*}}
\newcommand{\eneqs}{\end{equation*}}

\input{epsf}

\begin{document}

\tolerance 1000
				   
\title{ Heavy Fermion Quantum Criticality }

\author { Zaira Nazario$^\dagger$ and David I. Santiago$^\star$ }

\affiliation{$\dagger$ Max Planck Institute for the Physics of Complex
systems, N\"othnitzer Strasze 38, 01187 Dresden, Germany \\
$\star$ Instituut-Lorentz for Theortical Physics, Universiteit Leiden,
P. O. Box 9506, NL-2300 RA Leiden, The Netherlands}

\begin{abstract}

\begin{center}

\parbox{14cm}{ During the last few years, investigations of Rare-Earth
  materials have made clear that not only the heavy fermion phase in
  these systems provides interesting physics, but the quantum
  criticality where such a phase dies exhibits novel phase transition
  physics not fully understood. Moreover, attempts to study the
  critical point numerically face the infamous fermion sign problem,
  which limits their accuracy. Effective action techniques and
  Callan-Symanzik equations have been very popular in high energy
  physics, where they enjoy a good record of success. Yet, they have
  been little exploited for fermionic systems in condensed matter
  physics. In this work, we apply the RG effective action and
  Callan-Symanzik techiques to the heavy fermion problem. We write for
  the first time the effective action describing the low energy
  physics of the system. The $f$-fermions are replaced by a dynamical
  scalar field whose nonzero expected value corresponds to the heavy
  fermion phase. This removes the fermion sign problem, making the
  effective action amenable to numerical studies as the effective
  theory is bosonic. Renormalization group studies of the effective
  action can be performed to extract approximations to nonperturbative
  effects at the transition. By performing one-loop renormalizations,
  resummed via Callan-Symanzik methods, we describe the heavy fermion
  criticality and predict the heavy fermion critical dynamical
  susceptibility and critical specific heat. The specific heat
  coefficient exponent we obtain (0.39) is in excellent agreement with
  the experimental result at low temperatures (0.4).}

\end{center}

\end{abstract}


\date{\today}

\maketitle

For a couple of decades, heavy fermion materials have attracted the
focus of a large part of the experimental and theoretical condensed
matter community\cite{hfold, read}. There are many reasons for such a
spotlight on these materials. They exhibit exotic superconductivity,
interesting magnetism, but most importantly heavy quasiparticles with
an enlarged Fermi surface. This heavy quasiparticle phase perishes
into a quantum critical point with interesting, puzzling and not yet
understood nature\cite{si,hfcrit}.

It is quite striking to see the fermionic quasiparticle with masses
from tenths to about thousandths of an elementary electron mass. This
has been understood as arising from Kondo-like physics of the almost
localized $f-$electrons when they hybridize with the lower atomic
angular momentum bands of the material\cite{read, nagaosa}. This
hybridization gives rise to an enlarged Fermi surface as the
$f-$electrons now contribute to the Fermi volume, and to the large
quasiparticle mass and large specific heat coefficient, through
enhanced collective Kondo-like low energy scattering.  Of course, the
$f-$electrons have a strong tendency to localize due to their large
$U$ which fights the hybridization $V$.

Some of these materials can be tuned (by applying pressure, etc.)  so
that a critical value $(U/V)_c$ is reached such that, for values
larger than the critical value, the $f-$electrons localize and there
is no heavy fermion phase. Instead, there is a small Fermi surface
metal that usually exhibits magnetic order mediated via RKKY
interactions\cite{rkky}. For subcritical values of $U/V$, the system
is in the heavy fermion phase with large Fermi surface and no
magnetism.

At the critical value, a continous quantum phase transition occurs as
corroborated via scaling experiments. This transition is not
understood. The lack of understanding is a barrier to the full
characterization of the physical properties, phase diagram and
experimental response features of these materials. In this work, we
turn our attention to understanding this heavy fermion quantum
criticality\cite{si,coleman}.

We start from the partition function for $f-$electron hybridizing
with metallic $c-$electrons, i.e. the periodic Anderson Model
    \begin{widetext} 
      \begin{align}
	\begin{aligned} \nonumber
	  \mathcal Z &= \int \mathcal D c^\dagger \mathcal D c	 
          \mathcal D f^\dagger \mathcal D f \; e^{- \mathcal S}\\
	  \mathcal S &= \sum_M \int \frac{d \omega}{(2 \pi)} \frac{d^3
          \vec k}{(2 \pi)^3} \left[ - \omega \left( f_M^\dagger
          (\omega, \vec k) f_M (\omega, \vec k) + c_M^\dagger
          (\omega, \vec k) c_M (\omega, \vec k) \right) +
          \frac{(\epsilon_{\vec k} - \mu)}{\hbar} c^\dagger_M
          (\omega, \vec k) c_M (\omega, \vec k) \right] \\
        \end{aligned}
      \end{align}
      \begin{align}
        \begin{aligned}
	  &+ \sum_M \int \frac{d \omega}{(2 \pi)} \frac{d^3 \vec k}{(2
          \pi)^3} \frac{E_0-U}{\hbar} f_M^\dagger (\omega, \vec k) f_M
          (\omega, \vec k) - \frac{V}{\hbar} \sum_M \int \frac{d
          \omega}{(2 \pi)} \frac{d^3 \vec k}{(2 \pi)^3} \left(
          f_M^\dagger (\omega, \vec k) c_M (\omega, \vec k) +
          \text{h. c.} \right) \\ 
          &+ \frac{U}{\hbar} \sum_{M L} \int \frac{d \omega d \omega_2
          d \omega_3}{(2 \pi)^3} \frac{d^3 \vec k d^3 \vec k_2 d^3
          \vec k_3}{(2 \pi)^9} f_M^\dagger (\omega, \vec k) f_M
          (\omega_2, \vec k_2) f_L^\dagger (\omega_3, \vec k_3) f_L
          (\omega - \omega_2 + \omega_3, \vec k - \vec k_2 + \vec k_3)
	\end{aligned}
      \end{align}
    \end{widetext}
where the subscripts $M$, $L$ (and all capital letters subscripts for
that matter) indicate the angular momentum degeneracy of the ground
state ($j = 5/2$), $U$ is the Hubbard repulsion and $V$ represents the
strength of mixing of $f$-electrons with the conduction band.

If we write $n_i = \sum_M f_{i M}^\dagger f_{i M}$, the Hubbard interaction term 
then takes the form (in Euclidean time and real space) $
      \frac{U}{\hbar} \sum_i \int d \tau \left[ n_i^2 - n_i \right]$.
We can decouple this interaction term by using the
Hubbard-Stratonovich identity $
      e^{- \int d \tau \sum_i \frac{U}{\hbar} n_i^2} = \int \mathcal D
      \varphi e^{- \int d \tau \sum_i \frac{U}{\hbar} \left[
      \varphi_i^2 + 2 i \varphi_i n_i \right]}$ 
which in Euclidean time and momentum space reads $
      \int \mathcal D \varphi e^{- \int d \tau \frac{d^3 \vec k}{(2
      \pi)^3} \frac{U}{\hbar} \left[ \varphi_{\vec k} \varphi_{-\vec
      k} + 2 i n_{\vec k} \varphi_{- \vec k} \right]}$. 
Using $n_{\vec k} = \int \frac{d^3 \vec k}{(2 \pi)^3} f_{\vec
  q}^\dagger f_{\vec q - \vec k}$, this decoupling gives\cite{nagaosa}
    \begin{align}
      \begin{aligned}
        \nonumber
	&\mathcal Z = \int \mathcal D c^\dagger \mathcal D c \mathcal
	D f^\dagger \mathcal D f \mathcal D \varphi \; e^{- \mathcal
	S} \\ 
	&\mathcal S = \sum_M \int d\tau \frac{d^3 \vec k}{(2 \pi)^3}
	\left\{ f_{\vec k M}^\dagger \frac{\partial}{\partial \tau}
	f_{\vec k M} + \frac{E_0-U}{\hbar} f_{\vec k M}^\dagger f_{\vec
	k M} \right. \\
	&\left. + \frac{U}{N \hbar} \varphi_{\vec k} \varphi_{- \vec
	k} + c_{\vec k M}^\dagger \frac{\partial}{\partial \tau}
	c_{\vec k M} + \frac{(\epsilon_{\vec k} - \mu)}{\hbar}
	c^\dagger_{\vec k M} c_{\vec k M} \right\} \\
	&- \sum_M \int d \tau \frac{d^3 \vec k}{(2 \pi)^3}
	\frac{V}{\hbar} \left( f^\dagger_{\vec k M} \, c_{\vec k M} +
	c^\dagger_{\vec k M} f_{\vec k, M} \right) \\
	&+ 2 i \frac{U}{\hbar} \sum_M \int d \tau \frac{d^3 \vec k d^3
	\vec q}{(2 \pi)^6} \varphi_{- \vec k} f_{\vec q, M}^\dagger
	f_{\vec q - \vec k, M}
      \end{aligned}
    \end{align}
In order to have $f$ levels occupied we choose $E_0 < \mu$. 

This last action is what quantum field theorists would call the bare
action. It embodies the essential physics of the heavy fermion phase
and the heavy fermion criticality. Unfortunately, it is very hard to
solve the bare action exactly or numerically to high accuracy in order
to extract the desired information from it. The famous fermion sign
problem thwarts numerics, and exact solutions are normally impossible
in many body problems just as this one. Fortunately, there is a way
forward that can help extract some, and perhaps a lot of the
physics. The renormalization group\cite{wilson}, and in particular
effective action and Callan-Symanzik techniques\cite{wilzcek, gross,
  zj}, popular in particle physics, provide room for progress.

One of the lessons of the renormalization group is that as we
concentrate on longer wavelength, lower energy degrees of freedom, the
short distance and high energy fluctuations do two things. These
fluctuations renormalize the strength of the terms in the original
action and they generate new terms in the action which in turn change
as they get renormalized. The end result is that some terms in the
action become larger while others become smaller, thus not contributing
to the to the universal low energy physics of the system. We will thus
analyze the action for heavy fermion materials above, and obtain the
effective action with terms relevant to the low energy physics of the
heavy fermion phase and to the critical point where such a transition
perishes.

To obtain the effective action, rather than obtain the
renormalizations all at once, it proves advantageous to integrate out
the $f-$electrons and get those terms they contribute to the effective
action which are relevant for the low energy universal physics of the
critical point and heavy fermion phase. This is different from the
traditional approach where the $f-$electrons are integrated out and
all terms are kept, usually in the form of a functional determinant,
irrespective of whether the terms are relevant or not for the low
energy physics. The other important novelty in our analysis is the
application of RG effective action and Callan-Symanzik methods to
access the heavy fermion critical point from the effective action.

After the calculations are performed, the $f-$electrons disappear from
the theory. We have instead the Hubbard-Stratonovih field $\varphi$,
which has acquired dynamics through the $f-$electron
fluctuations. Such fluctuations also generate self-interaction terms
for the $\varphi$-field, and interaction terms between the $\varphi$'s
and the metallic $c-$electrons.

The effective action for the heavy fermion materials comes out to be
    \begin{widetext}
      \begin{align}
        \begin{aligned}
          \label{czaction}
          Z &= \int \mathcal D c_0^\dagger \mathcal D c_0 \mathcal D
          \tilde \varphi_0 \, e^{i S}\, , \text{ with }
          S = \sum_M \int \frac{d \omega}{2 \pi} \frac{d^3 \vec k}{(2
          \pi)^3} \left\{ \Lambda \, g_2 \, \tilde \varphi_0 (\omega,
          \vec k) \tilde \varphi_0 (- \omega, - \vec k) - \omega \,
          \tilde \varphi_0 (\omega, \vec k) \tilde \varphi_0 (-
          \omega, - \vec k) \right. \\
          &+ \left[ \omega - (\epsilon_{\vec k} - \mu) / \hbar \right]
          c_{0M}^\dagger (\omega, \vec k) c_{0M} (\omega, \vec k)
          - \Lambda^{-1/2} \, g_3 \, \int \frac{d \nu}{2 \pi}
          \frac{d^3 \vec q}{(2 \pi)^3} \, \tilde \varphi_0 (\omega,
          \vec k) c_{0M}^\dagger (\nu, \vec q) c_{0M} (\omega + \nu,
          \vec q + \vec k) \\
        \end{aligned}
      \end{align}
      \begin{align}
        \begin{aligned}
          \nonumber
          & \left. + \Lambda^{-2} \, g_4 \, \int \frac{d \omega_2 d
          \omega_3}{(2 \pi)^2} \frac{d^3 \vec k_2 d^3 \vec k_3}{(2
          \pi)^6} \, \tilde \varphi_0 (\omega, \vec k) \tilde
          \varphi_0 (\omega_2, \vec k_2) \tilde \varphi_0 (\omega_3,
          \vec k_3) \tilde \varphi_0 (- \omega - \omega_2 - \omega_3,
          -\vec k - \vec k_2 - \vec k_3) \right\}
        \end{aligned}
      \end{align}
    \end{widetext}
where
    \begin{align}
      \begin{aligned}
        \nonumber
        &\tilde \varphi_0 (\omega, \vec k) = \sqrt{\frac{4 U^2 (|E_0|
        + U)}{\pi D^3}} \, \varphi_0 (\omega, \vec k) \\
        &g_2 = \frac{D^2}{\hbar \Lambda (|E_0| + U)} - \frac{\pi
        D^3}{4 N U \Lambda \hbar (|E_0| + U)} 
      \end{aligned}
    \end{align}
    \begin{align}
      \begin{aligned}
        \nonumber
        g_3 = \sqrt{\frac{\pi D^3 V^4}{\hbar^2 \Lambda^2 (E_0 -
        U)^5}} \;; \quad
        g_4 = \frac{\pi D^3}{3 \hbar \Lambda (|E_0| + U)^2} \;.
      \end{aligned}
    \end{align}
We see that the universal physics of the heavy fermion system is
captured by an action of dynamical scalar fiels interacting with the
metallic $c-$electrons. We call this action the {\it heavy-fermion
  dynamic $\varphi^4$ action}. The heavy fermion phase corresponds to
$g_2$ being negative and $\varphi$ acquiring a nonzero expected value
as $\varphi$ is proportional to the density of $f-$electrons that
hybridizes with the metallic ones.  The heavy fermion critical point
occurs at $g_2 = 0$, when $\langle \varphi \rangle$ first becomes $0$.

The heavy-fermion dynamic $\varphi^4$ action is a new and important
result. It opens the door to accurate numerics for the fermion action,
as the interacting fermions that drive the transtion have been
replaced by a scalar field. This should eliminate the fermion sign
problem that plagues numerics, for all the action is happening in the
scalar fields and not the left-over metallic fermions.

One can apply standard order parameter RG to this action.
As an example, below we do a one-loop momentum shell renormalization
with the help of Callan-Symanzik equations {\it \`a la} 
Weinberg\cite{zj, weinberg, c-s}to resum and thus catch
some of the nonperturbative physics of the transition. This can of
course be improved by going to higher orders, and there is also plenty
of room to perform $\epsilon$-expansion studies instead of momentum
shell.

Since we can use the bare Fermi velocity of the metallic $c-$electrons
as a standard of speed in the material, we use it as such to express
our frequencies in units of momentum and work in ``God-given'' heavy
fermion units: $v_F=1$ and $\hbar=1$, $k_F=m=\Lambda$. After
renormalization, the heavy-fermion dynamic $\varphi^4$ action becomes
    \begin{widetext}
      \begin{align}
        \begin{aligned}
          Z &= \int \mathcal D c^\dagger \mathcal D c \mathcal D
          \tilde \varphi \, e^{i S} \, \text{ with }
          S = \sum_M \int \frac{d \omega}{2 \pi} \frac{d^3 \vec k}{(2
          \pi)^3} \left\{ \mu g_2^R \, Z_{\varphi}^2 \,
          \tilde \varphi (\omega, \vec k) \tilde \varphi (- \omega, -
          \vec k) - Z_{\varphi}^2 \omega \, \tilde \varphi (\omega,
          \vec k) \tilde \varphi (- \omega, - \vec k) \right. \\
          &+ Z_c^2 \left[ \omega - (\epsilon_{\vec k} - \mu)
          \right] c_M^\dagger (\omega, \vec k) c_M (\omega, \vec k) 
           - \mu^{-1/2} \, g_3^R  \, \int \frac{d
          \nu}{2 \pi} \frac{d^3 \vec q}{(2 \pi)^3} \, Z_{\varphi}
          Z_c^2 \, \tilde \varphi (\omega, \vec k) c_M^\dagger (\nu,
          \vec q) c_M (\omega + \nu, \vec q + \vec k) \\
          & \left. + \mu^{-2} \, g_4^R \, \int \frac{d \omega_2 d
          \omega_3}{(2 \pi)^2} \frac{d^3 \vec k_2 d^3 \vec k_3}{(2
          \pi)^6} Z_{\varphi}^4 \, \tilde \varphi (\omega, \vec k)
          \tilde \varphi (\omega_2, \vec k_2) \tilde \varphi
          (\omega_3, \vec k_3) \tilde \varphi (- \omega - \omega_2 -
          \omega_3, -\vec k - \vec k_2 - \vec k_3) \right\} \,.
        \end{aligned}
      \end{align}
    \end{widetext}

We now move to consider the specific momentum shell renormalizations
to determine the renormalization factors. The inverse $\varphi$
propagator, $\omega - \Lambda g_2$ goes into $- \mu g_2^R \,
Z_{\varphi}^2 \, + Z_{\varphi}^2 \omega + Z_k k_r = - \Lambda g_2 +
\omega + \Sigma_\varphi$, where
    \begin{align}
      \begin{aligned} \nonumber
        \Sigma_\varphi = \frac{g_3^2 \Lambda}{8 \pi^4 } \, (1 - \mu /
        \Lambda) \left [ 1 - \frac{(\omega - k_r)}{4 \Lambda} \right]
        + \frac{g_4 \, \, \Lambda}{2 \pi} \frac{(1 - \mu /
        \Lambda)}{(g_2 - 1)}
      \end{aligned}
    \end{align}

Next we tackle the renormalization of the $c-$ electron inverse
propagator $ \omega - (\epsilon_{\vec k} - \mu) $ which becomes $Z_c^2
\left[ \omega - (\epsilon_{\vec k} - \mu)  \right] = \omega -
(\epsilon_{\vec k} - \mu)  + \Sigma_c (\omega, \vec k)$, with
    \begin{align}
      \begin{aligned}
        \nonumber
	&\Sigma_c (\omega, \vec k) = - \frac{g_3^2 \, \Lambda }{(2 \pi)^4}
        \, (1 - \mu / \Lambda) \left\{ \left[ \frac{1}{  (g_2 -
        1)} \right] \ln \left| \frac{g_2 + 1}{g_2 - 1} \right|
        \right. \\
      \end{aligned}
    \end{align}
    \begin{align}
      \begin{aligned}
        &- \frac{4 g_2}{\Lambda \left( g_2^2 - 1 \right)^2} (\omega
        - k_r) \ln \left| \frac{\omega - k_r}{2 \Lambda} \right| \\
        &+ \left. (\omega - k_r) \left[ \frac{2 g_2}{ \Lambda (g_2^2
        - 1)} - \frac{1}{\Lambda (g_2 - 1)^2} \ln \left|
        \frac{g_2 + 1}{g_2 - 1} \right| \right] \right\}
      \end{aligned}
    \end{align}
Notice that the $c-$electron propagator only renormalizes trivially to
this order at the critical point. The reason is that if $g_2$ is
nonzero, we need to include the nonzero expected value of $\varphi$
and expand around it in the heavy fermion phase in order to get the
physics right. We are not interested in the heavy phase but in the
critical point, where such a phase disappears. At the critical point,
$g_2 = 0$ and things work out as expected.

We now turn our attention to vertex renormalizations, starting with
the $\varphi-c$ vertex, $\Lambda^{1/2} \, g_3$. It renormalizes into
$- \mu^{- 1/2} \, g_3^R \, Z_{\varphi} Z_c^2 = - \Lambda^{- 1/2} \,
(g_3 + \Gamma_c)$, where
    \begin{align}
      \begin{aligned} \nonumber
	\Gamma_c &= \frac{g_3^2 }{(2 \pi)^4} (1 - \mu /
        \Lambda) \left\{ \frac{4 g_2}{ (g_2^4 - 1)} -
        \frac{2 \pi g_2}{ (g_2^2 - 1)^2} \right. \\
        &+ \left. \frac{4 (g_2^2 + 1)}{ (g_2^2 - 1)^2}
        \arctan \left( \frac{1}{g_2} \right) \right\}
      \end{aligned}
    \end{align}
And finally to the renormalization of the $\varphi^4$ interaction,
$\mu^{-2} \, g_4^R \, Z_\varphi^4 = \Lambda^{-2} g_4 + \Gamma_\varphi$,
with
    \begin{align}
      \begin{aligned} \nonumber
        \Gamma_\varphi = - \frac{g_3^4 }{24 \pi^4 \Lambda^2} \,
        (1 - \mu / \Lambda) + \dots
      \end{aligned}
    \end{align}
    \begin{align}
      \begin{aligned} \nonumber
        \dots  \frac{g_4^2 }{(2 \pi)^4 \Lambda^2} (1 - \mu / \Lambda)
        \left\{ \left[ \frac{1}{g_2 + 1} \right]^2 + \left[
        \frac{1}{g_2 - 1} \right]^2 \right\}
      \end{aligned}
    \end{align}

The heavy fermion criticality occurs when $g_2=0$. In the $g_2=0$
critical manifold
    \begin{align}
      \begin{aligned} \nonumber
	&Z_c = 1 \, , \, \, \,
        Z_\varphi \simeq 1 - \frac{g_3^2 }{64 \pi^4 
        } \, (1 - \mu / \Lambda) \, , \, \, \,
        Z_k = \frac{g_3^2 }{32 \pi^4  } \\
        & \frac{g_3^R}{\mu^{1/2}} \simeq \frac{g_3}{\Lambda^{1/2}} \left\{ 1
        +  \frac{g_3 }{(2 \pi)^3 } \left[
        \frac{g_3}{8 \pi} - 1 \right] (1 - \mu /
        \Lambda) \right\} \\
        &\frac{g_4^R}{\mu^{2}} \simeq \frac{g_4}{\Lambda^{2}} + \left[ \frac{2
        g_4^2}{(2 \pi)^4  \Lambda^2} - \frac{g_3^4 m^2}{24
        \pi^4  \Lambda^4} - \frac{g_3^2 g_4}{(2 \pi)^4 
        \Lambda^2} \right] (1 - \mu / \Lambda) \,.
      \end{aligned}
    \end{align}
To this order of approximation
    \begin{align}
      \begin{aligned}
        \nonumber
        \beta_3 &= \mu \frac{\partial g_3^R}{\partial \mu} \Big |_{\mu
        = \Lambda} = \frac{1}{2} g_3 - \frac{g_3^2}{(2 \pi)^3
         } \left[ \frac{g_3}{8 \pi} - 1
        \right] \\
        g_3^c &= 4 \pi \left[ 1 + \sqrt{1 + 2 \pi^2}
        \right] \\
        \beta_4 &= \mu \frac{\partial g_4^R}{\partial \mu} \Big |_{\mu
        = \Lambda} \!\!\!\!\!\!\! = 2 g_4 - \frac{6 g_4^2}{(2 \pi)^4
        } + \frac{g_3^4 }{8 \pi^4} + \frac{3
        g_3^2 g_4 }{(2 \pi)^4} \\
        g_4^c &= \frac{g_3^{c2}}{4} + \frac{(2 \pi)^4}{6} +
        \sqrt{\frac{33 g_3^{c4}}{16 } + \frac{(2 \pi)^8
        }{36} + \frac{(2 \pi)^4}{12} g_3^{c2}}
      \end{aligned}
    \end{align}
The $\varphi$ anomalous dimension at cricitcality is
    \begin{align}
      \begin{aligned}
        \nonumber
        \gamma_\varphi = \mu \frac{\partial \ln Z_\varphi}{\partial
        \mu} \Big |_{\mu = \Lambda} = \frac{g_3^2}{64 \pi^4 } = \frac{1}{2 \pi^2 } \left[ 1 + \pi^2 
      + \sqrt{1 + 2 \pi^2 } \right] \, .
    \end{aligned}
    \end{align}

We can make experimental predictions of our heavy fermion $\varphi^4$
action. These predictions are part of our new results and serve as an
experimental test of the validity of our theory. Among those are the
susceptibility and specific heat of the system. In the heavy fermion
phase, the $f$-electrons contribute to the susceptibility and specific
heat since they are mixed with the metallic band electrons, forming in
conjunction one ``Fermi liquid''. The susceptibility of conduction
electrons is negligible since they are nonmagnetic. The susceptibility
is proportional to the number of $f$-electrons being ``pulled'' by an
applied field, and hence to their density of states, which is given by
the imaginary part of the $f-$electron propagator\cite{read}. Such a
propagator is proportional to $\langle \varphi \rangle$, which
satisfies a Callan-Symanzik equation\cite{zj,weinberg}
    \begin{align}
      \begin{aligned}
        \nonumber
        0 &= \left[ \mu \frac{\partial}{\partial \mu} +
        \frac{\gamma_\varphi}{2} \right] G^{(1)} \Big |_{\mu = \omega} \\
        G^{(1)} (\omega) &= \left( \frac{1}{\omega} \right)^{\gamma_\varphi /
        2} \sim  \left( \frac{1}{\omega}
        \right)^{[ 1 + \pi^2 + \sqrt{1 + 2 \pi^2} ] / (4 \pi^2)}\\
        F (\omega, \vec k = 0) &\sim i \langle \varphi (\omega, \vec k =
        0) \rangle = i G^{(1)} (\omega) 
      \end{aligned}
    \end{align}
    \begin{align}
      \begin{aligned}
        \chi (\omega) & = \lim_{\epsilon \rightarrow 0} Im \left[
        F(\omega + i \epsilon) \right] \sim 
      \left( \frac{1}{\omega} \right)^{\gamma_\varphi /
        2} \, .
      \end{aligned}
    \end{align}

The specific heat coefficient is also proportional to the density of
states, which is inversely proportional to the Fermi energy\cite{read}. The
$f$-electrons, being quasi-localized, form a quite thin band and hence
have a small $E_F$. Thus their density of states is so big in
comparison with that of the conduction electrons, that the density of
states of the ``mixed Fermi liquid'' can be approximated to be that of
the $f$-electrons. We obtain
    \begin{align}
      \begin{aligned}
        C_T &\sim T \, \lim_{\epsilon \rightarrow 0} Im \left[
        F(\omega + i \epsilon) \right] \Big |_{\omega \sim T} 
      \sim T \left( \frac{1}{T} \right)^{\gamma_\varphi / 2}
      \end{aligned}
    \end{align}
We obtain a specific heat coefficient exponent $\gamma_\varphi / 2 =
0.39$. This is in excellent agreement with the exponent $0.4$ found
for YbRh$_2$Si$_2$ at low temperatures\cite{si}.

Via renormalization group studies and effective action techniques
common to field theories of particle physics, we have obtained the
effective field theory for heavy fermion quantum criticality. This
marks important progress as the effective field theory is bosonic,
vitiating the fermion sign problem and thus being amenable to
numerical studies and high order $\epsilon$ expansion studies. 

The critical field theory can be studied using the renormalization
group. We did so via one-loop renormalization studies, improved by
means of Callan-Symanzik resumations to access some of the
nonpertrubative effects. We thus make predictions for the exponents
that characterize the critical divergence of the specific heat
coefficient and the critical charge susceptibility. Our specific heat
coefficient exponent of 0.39 is in excellent agreement with the 0.4
found in experiments at low temperatures\cite{si}.

\end{document}